\documentclass[10pt,twocolumn,letterpaper]{article}

\usepackage{cvpr,times,epsfig,graphicx,amsmath,amssymb,booktabs,makecell}
\newlength\savewidth
\usepackage[pagebackref=true,breaklinks=true,letterpaper=true,colorlinks,bookmarks=false]{hyperref}

 \cvprfinalcopy 
\def\cvprPaperID{****}

\ifcvprfinal\fi
\begin{document}


\title{“Predicting Global Variations in Outdoor PM$_{2.5}$ Concentrations using Satellite Images and Deep Convolutional Neural Networks
}
\author{Kris Y. Hong\\
McGill University\\
Montreal, QC, Canada\\
\and
Pedro O. Pinheiro\\
Element AI\\
Montreal, QC, Canada\\
\and
Scott Weichenthal\\
McGill University\\
Montreal, QC, Canada\\
scott.weichenthal@mcgill.ca
}

\maketitle

\begin{abstract}
Here we present a new method of estimating global variations in outdoor PM$_{2.5}$ concentrations using satellite images combined with ground-level measurements and deep convolutional neural networks. Specifically, new deep learning models were trained over the global PM$_{2.5}$ concentration range ($<$1-436 $\mu$g/m$^3$) using a large database of satellite images paired with ground level PM$_{2.5}$ measurements available from the World Health Organization. Final model selection was based on a systematic evaluation of well-known architectures for the convolutional base including InceptionV3, Xception, and VGG16. The Xception architecture performed best and the final global model had a root mean square error (RMSE) value of 13.01 $\mu$g/m$^3$ (R$^2$=0.75) in the disjoint test set. The predictive performance of our new global model (called IMAGE-PM$_{2.5}$) is similar to the current state-of-the-art model used in the Global Burden of Disease study but relies only on satellite images as input.  As a result, the IMAGE-PM$_{2.5}$ model offers a fast, cost-effective means of estimating global variations in long-term average PM$_{2.5}$ concentrations and may be particularly useful for regions without ground monitoring data or detailed emissions inventories. The IMAGE-PM$_{2.5}$ model can be used as a stand-alone method of global exposure estimation or incorporated into more complex hierarchical model structures.
\end{abstract}

\section{Introduction}\label{sec:intro}
Environmental pollution is a global health concern with economic impacts measured in billions of dollars each year~\cite{landrigan2018lancet}. In particular, ambient fine particulate air pollution (PM$_{2.5}$) kills millions of people around the world annually and is consistently ranked among the leading global burden of disease risk factors~\cite{stanaway2018global}. 

In recent years, incredible progress has been made in estimating global variations in outdoor PM$_{2.5}$ concentrations through the combined use of multiple complex data streams including remote sensing estimates of aerosol optical depth, chemical transport models, and ground-level geographic information~\cite{brauer2012exposure,van2016global,shaddick2018data}. 

Other approaches to air pollution exposure assessment include the use of statistical models (\eg land use regression models) that combine geographic information system (GIS) data with ground monitoring data to predict exposures in locations without measurements. While this approach generally works well~\cite{weichenthal2016characterizing,ryan2007review}, detailed GIS data are often available on a limited spatial scale and land use regression models are not generalizable across cities~\cite{patton2015transferability}. Alternatively, information on traffic, land use, the built environment, and other potential sources of exposure can also be captured in digital images both locally and through satellite imagery.

As such, large databases of paired pollutant-image samples may provide an alternative, cost-effective means of training deep convolutional neural networks~\cite{lecun1998gradient} for the purpose of estimating environmental exposures across broad geographic areas including regional variations in ambient PM$_{2.5}$ concentrations~\cite{weichenthal2019picture}.

\begin{figure}[!t]
\begin{center}
\includegraphics[width=1\linewidth]{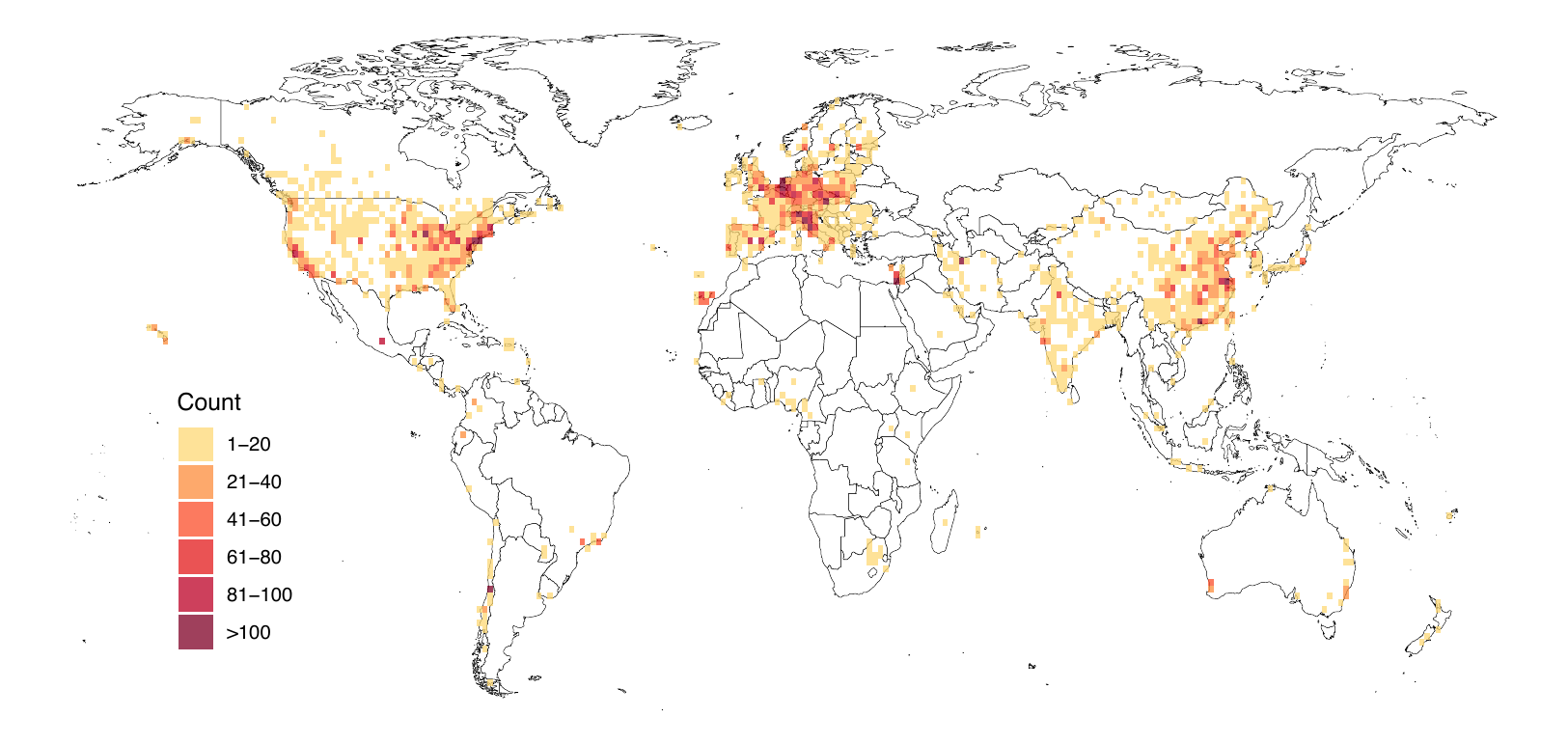}
\end{center}
\caption{Locations of global monitoring sites for PM$_{2.5}$.}
\label{fig:monitor_location}
\end{figure}

While deep learning image analysis is increasingly used for computer vision applications in medicine~\cite{esteva2017dermatologist,gulshan2016development,cruz2017accurate,angermueller2016deep}, little work has focused on combining digital images with deep convolutional neural networks for the purpose of estimating environmental exposures~\cite{maharana2018use}. 
Nevertheless, recent applications of deep learning in environmental health research have provided encouraging results including reliable predictions of spatial differences in obesity prevalence based on built environment characteristics~\cite{maharana2018use}. 

In this study, our goal was to explore the use of deep convolutional neural networks in estimating global variations in annual average outdoor PM$_{2.5}$ concentrations using only satellite images. Specifically, we examined the performance of a series of deep convolutional neural networks in estimating outdoor PM$_{2.5}$ concentrations across the global exposure range as well as over the more limited exposure range of North America. While the global PM$_{2.5}$ database covered the entire North American exposure range, a separate North American model was developed to examine the applicability of this method across a narrow concentration gradient.

\section{Method}\label{sec:method}
\subsection{Long-Term Average Outdoor PM$_{2.5}$ Data}
A global database of annual average ground-level PM$_{2.5}$ measurements and corresponding latitude-longitude coordinates was compiled from the World Health Organization~\cite{who16data}. 

These data were collected primarily between 2010 and 2016 (89 samples were collected in 2017 and 142 samples were collected between 2000 and 2009) and included approximately 20,000 measurements from approximately 6,000 unique monitoring sites in 98 countries. We did not include PM$_{2.5}$ measurements estimated from PM$_{10}$ values in training our models. The locations of global monitoring sites are shown in  Figure~\ref{fig:monitor_location}.

In North America, a database of long-term average outdoor PM$_{2.5}$ concentrations (2010-2012) was obtained at a 0.01 decimal degree grid resolution (approximately 1km apart) from the Atmospheric Composition Analysis Group at Dalhousie University, Canada~\cite{atmcomp}. These data were estimated by combining aerosol optical depth information with the GEOS-Chem chemical transport model with subsequent calibration to ground-based observations using geographically-weighted regression~\cite{van2016global,atmcomp}. 
In total, the North American database included between approximately 87,000 and 623,000 ground level PM$_{2.5}$ estimates depending on the zoom level used for satellite images (described below). 

\subsection{Satellite Images}
Satellite images centred on each latitude-longitude pair for ground-level PM$_{2.5}$ data were downloaded from Google Static Maps using the ggmap package in the R statistical computing environment~\cite{Rlang,ggmap}. Four satellite images were downloaded for each monitoring site in the global database, differing by integer zoom levels ranging from 13 (covering approximately $10\times10km$) to 16 (approximately $1.5\times1.5km$). All images were saved at a resolution of $256\times256\times3$ to maintain a reasonable training time; zoom level 16 was excluded from the North American database owing to the excessive training times required. All satellite images were dated between September-December 2018. 

\subsection{Data Processing}
All latitude-longitude coordinates for PM$_{2.5}$-image pairs in the global database were first geohashed to a precision of three~\cite{geohash}. The geohashing process maps each latitude-longitude pair onto a global grid of rectangular cells, where each cell is defined by a unique geohash code. The resolution of the global grid depends on the precision level: the selected precision level of three corresponds to cells with areas less than approximately $156\times156km$, with widths decreasing moving from the equator to the poles.  The global database was then randomly split into training (80\%), validation (10\%), and test sets (10\%) such that the three sets were disjoint by geohash codes. This ensured that satellite images were disjoint between sets, allowing us to evaluate the generalizability of model estimates to data not encountered during the training process. 

During the model development phase, the training set was used to fit model weights, and the validation set was used for hyperparameter tuning (\ie choosing the optimal convolutional base, optimizer, and learning rate; and also adjusting the learning rate between epochs and deciding when to cease training using callbacks as described below). As both the training and validation sets were used to build and select the final model, an independent test set (which played no role in model training or selection) was used to evaluate final model performance. 

Multiple ground-level measurements were available for annual average PM$_{2.5}$ concentrations for some sites in the global database. This meant that multiple exposure values (\ie year-to-year changes in annual average PM$_{2.5}$ concentration at the same location over time) could be assigned to the same satellite image. We approached this issue in two ways: 1) Models were developed averaging all available exposure data for each latitude-longitude pair; 2) Models were developed without averaging allowing individual images to have different exposure values based on changes in annual average PM$_{2.5}$ concentrations over time. Preliminary models favoured the second approach (\ie allowing the same image to have different PM$_{2.5}$ concentrations over different years) and therefore only the second approach was explored in detail. As a result, global model evaluation was also based on single year annual average ground-level measurements. 

For the North American database, we extracted PM$_{2.5}$-image pairs from the complete dataset at decimal degree resolutions of 0.15, 0.10, and 0.05, for zoom levels 13, 14, and 15, respectively. These resolutions were selected such that satellite images did not overlap in area within each zoom level. These data were then randomly split into training, validation, and test sets. 

\subsection{Model Training and Evaluation}
Models were developed to predict spatial variations in outdoor PM$_{2.5}$ concentrations on a continuous scale using linear activations as well as across deciles of exposure (ten ordinal categories of exposure split by deciles) using \emph{softmax}~\cite{bridle90softmax} activations. All models included a convolutional base for feature extraction with an input size of $256\times256\times3$ (\ie width x height x color channels). Dropout layers with rates of 0.5 were included after the convolutional base and after the densely connected network to minimize overfitting. ImageNet weights were used for model initialization, and all models were trained using a batch size of 64 images (16 images per GPU) for up to 100 epochs. During model training, callback functions were used to: 1) Decrease the learning rate by a factor of 0.1 if the validation accuracy did not improve for 10 epochs; and 2) Stop model training if the validation accuracy did not improve for 20 epochs. For each of the four tasks of predicting continuous/categorical PM$_{2.5}$ on the global/North American scales, the model with the highest validation classification accuracy (for decile predictions) or the lowest validation root mean square error (RMSE) (for continuous predictions) was retained. For categorical models, we also report the ``one-off accuracy'' which reflects the proportion of the time the model predicts the correct class or one category away from the correct class (\eg predicting decile 9 when the true value is decile 10).

Final model selection was based on a systematic evaluation of several well-known architectures for the convolutional base including InceptionV3~\cite{szegedy2016rethinking}, Xception~\cite{chollet2017xception}, and VGG16~\cite{simonyan2014vgg}. In addition, several optimizers were tested including RMSProp~\cite{tieleman2012rmsprop} and Nadam~\cite{dozat2016nadam} with learning rates of 0.001 and 0.0001. A detailed leaderboard was maintained, tracking the performance of different combinations of model architectures and hyper-parameters; the model that performed best on the validation dataset was selected as the final model. 

Generally, the InceptionV3 and Xception architectures combined with the Nadam optimizer at a learning rate of 0.0001 performed best on the data, and these results are described in detail. For the final models, gradient-weighted class activation maps~\cite{selvaraju2017grad} and filter visualizations were used to examine specific portions of images used to make predictions and to evaluate which features were learned at various layers of the model. All analyses were conducted using the Keras package~\cite{keras} in R and Python with two Lambda Quad Workstations (Lambda Labs, San Fransisco, CA) containing 4 GPUs each (NVIDIA Titan Xp or 1080 Ti). On average, global model training took 2-minutes/epoch whereas the North America model took 10, 20, or 60-minutes/epoch for zoom levels 13, 14, and 15, respectively. 

As an additional model evaluation step, we compared continuous PM$_{2.5}$ estimates from our final global model (called IMAGE-PM$_{2.5}$) to those of the Data Integration Model for Air Quality (DIMAQ) used by the Global Burden of Disease study~\cite{shaddick2018data,shaddick2018datab}. 

This comparison was conducted for approximately 9000 locations (113 countries) between 2010 and 2016 (approximately 4000-6000 measurements per year) with 34,794 annual average measurements ranging from $<$1 $\mu$g/m$^3$ to 332 $\mu$g/m$^3$ (mean=20.04 $\mu$g/m3, SD=18.76 $\mu$g/m$^3$). In addition, we compared our global model estimates to mean DIMAQ estimates averaged over the entire 2010-2016 period. Finally, we calculated site-specific differences between our IMAGE-PM$_{2.5}$ estimates and mean DIMAQ estimates (2010-2016) to evaluate potential geographic patterns in the magnitude of disagreement between the two models.

\subsection{Data Availability}
All PM$_{2.5}$ data, code, image files, and final deep learning models are freely available upon request.

\section{Results}\label{sec:results}
\begin{table*}[!t]
\centering
\resizebox{2\columnwidth}{!}{%
\begin{tabular}{lccccccccccccccc}
\toprule
Database & Zoom & n & Mean & SD & \multicolumn{11}{c}{Decile}\\
\cmidrule{6-16}
&&&&& Min & 1 & 2 & 3 & 4 & 5 & 6 & 7 &8 & 9 & Max \\
\midrule
Global & 13-16 & 19,657 & 23.24 & 22.94 & 0.50 & 7.00 & 8.49 & 9.79 & 11.51 & 14.03 & 17.64 & 24.05 & 35.78 & 54.81 & 436.44 \\
& 13 & 87,104 & 3.98 & 2.23 & 0.00 & 1.90 & 2.30 & 2.70 & 3.00 & 3.40 & 3.80 & 4.40 & 5.30 & 7.80 & 14.00 \\
N. America & 14 & 194,739 & 3.98 & 2.23 & 0.00 & 1.90 & 2.30 & 2.70 & 3.00 & 3.40 & 3.80 & 4.40 & 5.30 & 7.80 & 15.50 \\
& 15 & 623,759 & 4.36 & 2.30 & 0.00 & 2.10 & 2.60 & 3.00 & 3.30 & 3.70 & 4.20 & 4.80 & 6.10 & 8.30 & 16.40 \\
\bottomrule
\end{tabular}
}
\vspace{1mm}
\caption{Descriptive statistics for the PM$_{2.5}$ ($\mu$g/m$^3$) data in the Global and North American databases.}
\label{tab:dataset_stats}
\end{table*}

The global database contained approximately 19,650 pollution-image pairs with annual mean PM$_{2.5}$ concentrations ranging from less than 1 $\mu$g/m$^3$ to 436 $\mu$g/m$^3$ with a mean value of 23.2 $\mu$g/m3 (SD= 22.9 $\mu$g/m$^3$) (Table~\ref{tab:dataset_stats}). Estimated PM$_{2.5}$ concentrations were much lower for North American ranging from less than 1 $\mu$g/m$^3$ to 16.4 $\mu$g/m$^3$ with a mean value of 4.36 $\mu$g/m$^3$ (SD=2.30).

\begin{table}[!t]
\centering
\resizebox{1\columnwidth}{!}{%
\begin{tabular}{ccccc}
\toprule
\makecell{Model\\Architecture} & Zoom & \makecell{Decile Class.\\Accuracy (\%)} & \makecell{SD\\(PM$_{2.5}$)} & RMSE \\
\midrule
Global Model & & & & \\
InceptionV3 & 13 & 32.38 & 23.70 & 13.87 \\
            & 14 & 30.16 & 23.70 & 13.86 \\
            & 15 & 30.32 & 23.70 & 14.03 \\
            & 16 & 30.00 & 23.70 & 14.22 \\
\hline
Xception    & 13 & 35.33 & 23.70 & 13.63 \\
            & 14 & 33.06 & 23.70 & 14.18 \\
            & 15 & 31.61 & 23.70 & 13.64 \\
            & 16 & 31.61 & 23.70 & 14.31 \\
\midrule
North American Model & & & & \\
InceptionV3 & 13 & 42.28 & 2.21 & 0.85 \\
            & 14 & 43.81 & 2.24 & 0.83 \\
            & 15 & 48.88 & 2.31 & 0.77 \\
\hline
Xception    & 13 & 44.66 & 2.21 & 0.77 \\
            & 14 & 45.95 & 2.24 & 0.74 \\
            & 15 & 50.95 & 2.31 & 0.72 \\
\bottomrule
\end{tabular}
}
\vspace{1mm}
\caption{Model performance on the validation set across different model architectures and zoom levels. The standard deviation of PM$_{2.5}$ values in the validation set are shown as a baseline for evaluating RMSE values.}
\label{tab:res}
\end{table}

In models classifying PM$_{2.5}$ concentrations across deciles, the Xception model architecture performed best in both the global and North American databases (Table~\ref{tab:res}). Specifically, the final global categorical model (using the Xception base and zoom level-13 for satellite images) had a validation accuracy of 35.33\% across deciles (10\% accuracy would be expected by random chance) (Table~\ref{tab:res}). 

\begin{figure}[t]
\begin{center}
\includegraphics[width=1\linewidth]{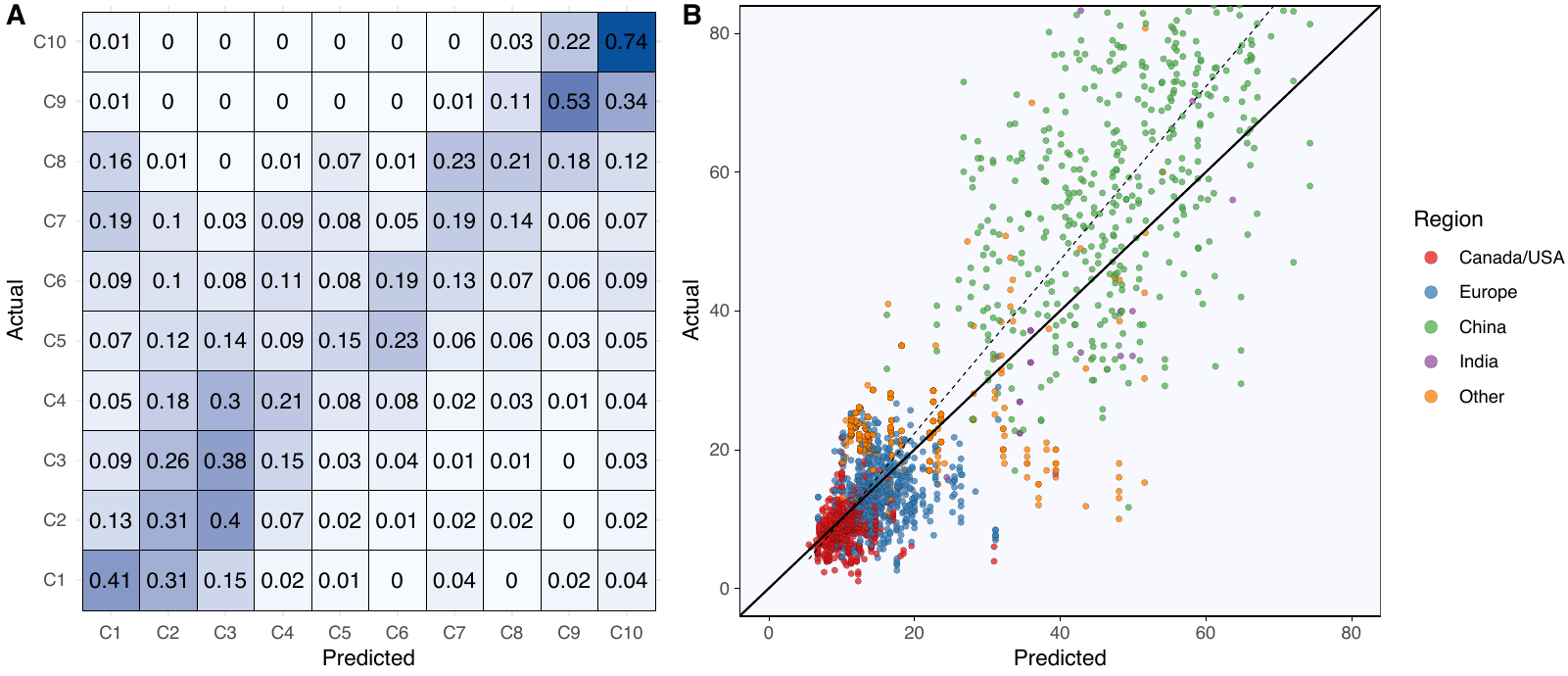}
\end{center}
\caption{Measured versus predicted global PM$_{2.5}$ concentrations in the test set for 10-category classification (A) and regression (B). The final model uses the Xception base with zoom level-13 satellite images.}
\label{fig:res_global}
\end{figure}

\begin{figure}[t]
\begin{center}
\includegraphics[width=1\linewidth]{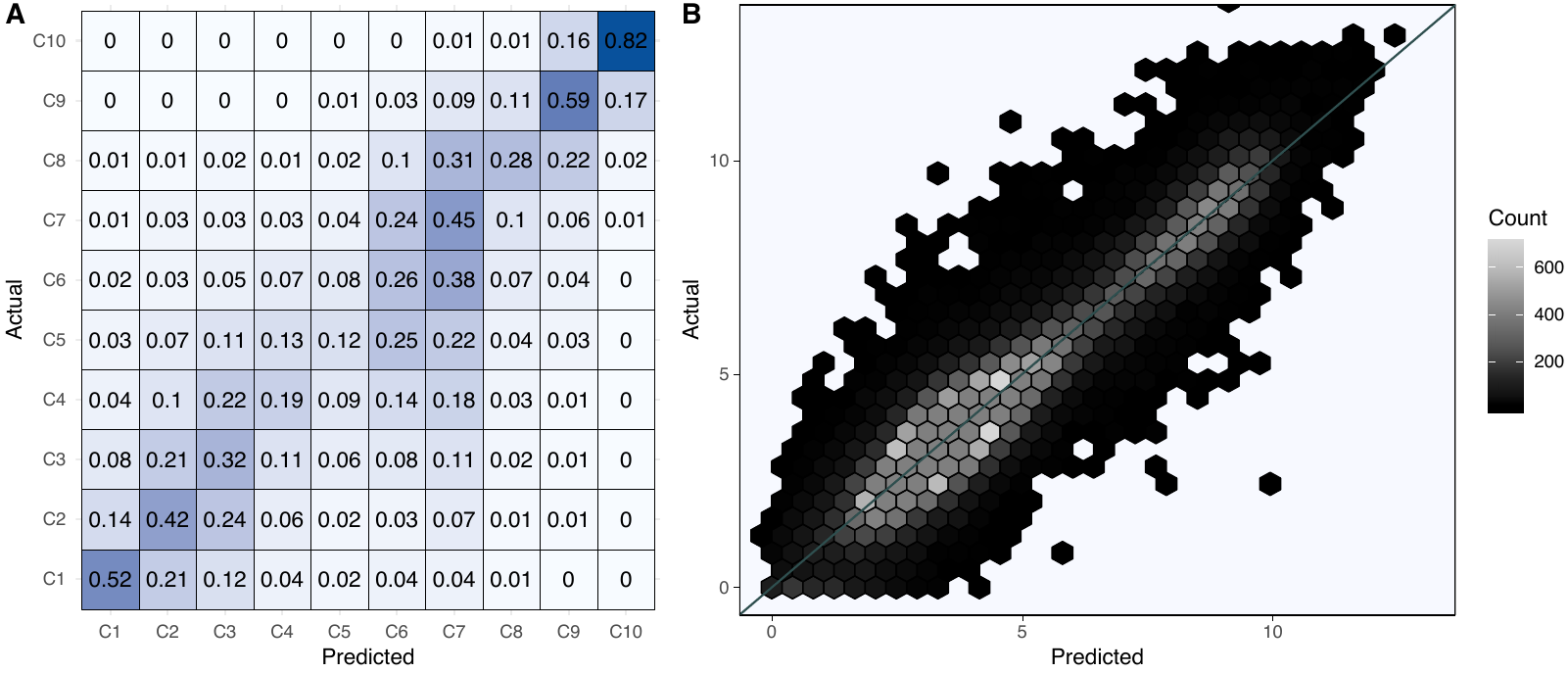}
\end{center}
\caption{Measured versus predicted North American PM$_{2.5}$ concentrations in the test set for 10-category classification (A) and regression (B). The final model uses the Xception base with zoom level-15 satellite images.}
\label{fig:res_na}
\end{figure}

The confusion matrix presented in Figure~\ref{fig:res_global}A illustrates model performance on the test set and indicates that categorical predictions were best at lower and upper deciles with decreasing performance towards the inner classes. Overall, the global categorical model achieved a test accuracy of 33.69\% and a one-off test accuracy of 65.71\%.  The final categorical model for North America (using the Xception base and zoom level-15 satellite images) achieved a validation accuracy of 50.95\% (Table~\ref{tab:res}). As with the global categorical model, the North America model performed better at the extremes (Figure~\ref{fig:res_na}A) with poorer accuracy for the central classes. The test accuracy of the model was 47.07\% and its one-off test accuracy was 78.41\%.

The Xception model architecture also performed best for continuous models in both the global and North American databases. For the global IMAGE-PM$_{2.5}$ model (using the Xception base model and zoom level-13 satellite images), the lowest validation RMSE value was 13.63 $\mu$g/m$^3$ (Table ~\ref{tab:res}). On the test dataset, the global model achieved an RMSE value of 13.01 $\mu$g/m$^3$ with an R$^2$ value of 0.75 (Figure 1~\ref{fig:res_global}B); however, model predictions tended to underestimate measured values at higher concentrations as indicated by the dashed fit-line in Figure~\ref{fig:res_global}B. In North America, the best continuous model (using the Xception base and zoom level-15 satellite images) had a validation RMSE of 0.72 $\mu$g/m$^3$ (Table~\ref{tab:res}). This model achieved an RMSE of 0.74 $ \mu$g/m$^3$ on the test set with an R$^2$ value of 0.89. A plot of measured versus predicted values in the test set is shown for North American in Figure~\ref{fig:res_na}B with the predictions generally following the 1:1 line. 

\begin{figure}[t]
\begin{center}
\includegraphics[width=1\linewidth]{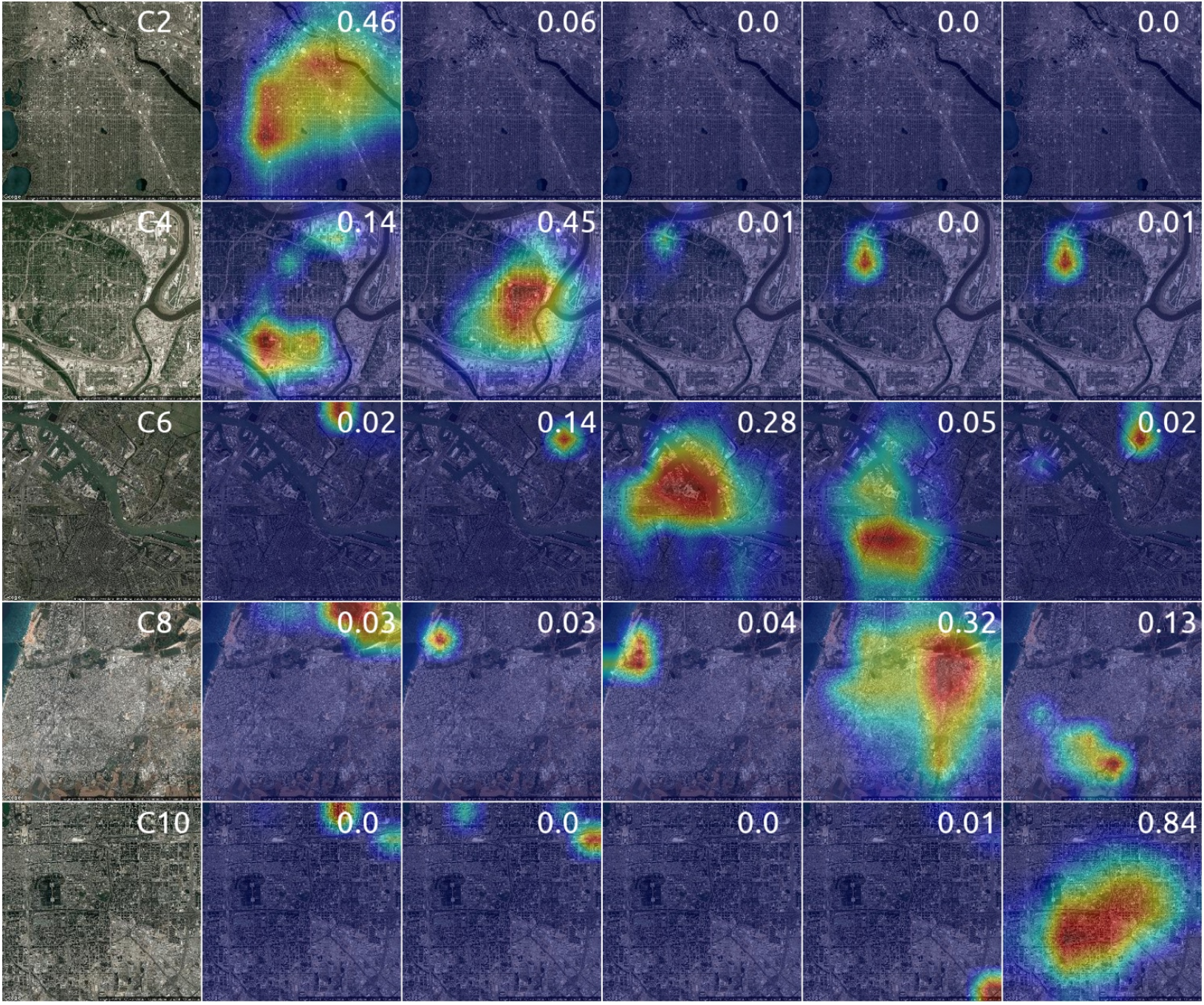}
\end{center}
\caption{Gradient-weighted class activation maps (Grad-CAMs) for images correctly classified by the final global categorical model (using the Xception base and zoom level-13 satellite images). The first column is the original input image. The second through sixth columns are the Grad-
CAMs for classes 2, 4, 6, 8, and 10, respectively. Numerical values on the top-right indicate the predicted probability that the image belongs to the respective class. The cities are Minneapolis, US (C2); Kansas City, US (C4); Amsterdam, NL (C6); Tel Aviv, IL (C8); and Beijing, CN (C10).}
\label{fig:res_qualitative}
\end{figure}


Gradient-weighted class activation maps and filter visualizations were used to identify specific portions of images used for predictions and to examine patterns learned by models in convolution layers, respectively. Class-activation maps are presented in Figure~\ref{fig:res_qualitative} for five locations that were correctly classified across deciles of long-term PM$_{2.5}$ concentrations. From this figure it is clear that localized portions of each satellite image are generally being used to make predictions; however, the specific ground-level features that are playing the most important role remain unclear. 

\begin{figure}
\begin{center}
\includegraphics[width=1\linewidth]{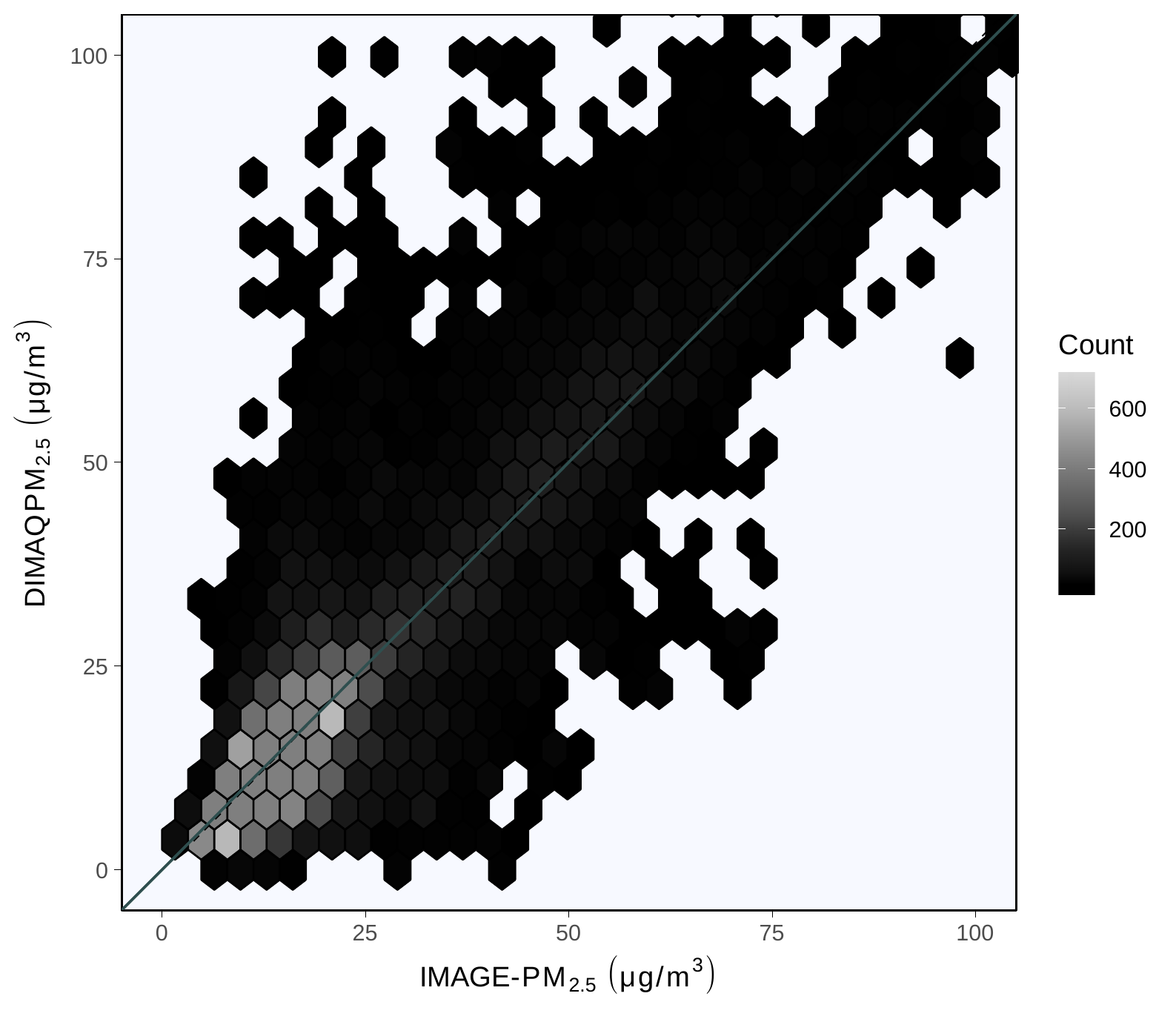}
\end{center}
\caption{Relationship between annual average PM$_{2.5}$ concentrations predicted by DIMAQ PM$_{2.5}$ and IMAGE-PM$_{2.5}$}
\label{fig:comp_dimaq}
\end{figure}

\begin{figure*}[t]
\begin{center}
\includegraphics[width=1\linewidth]{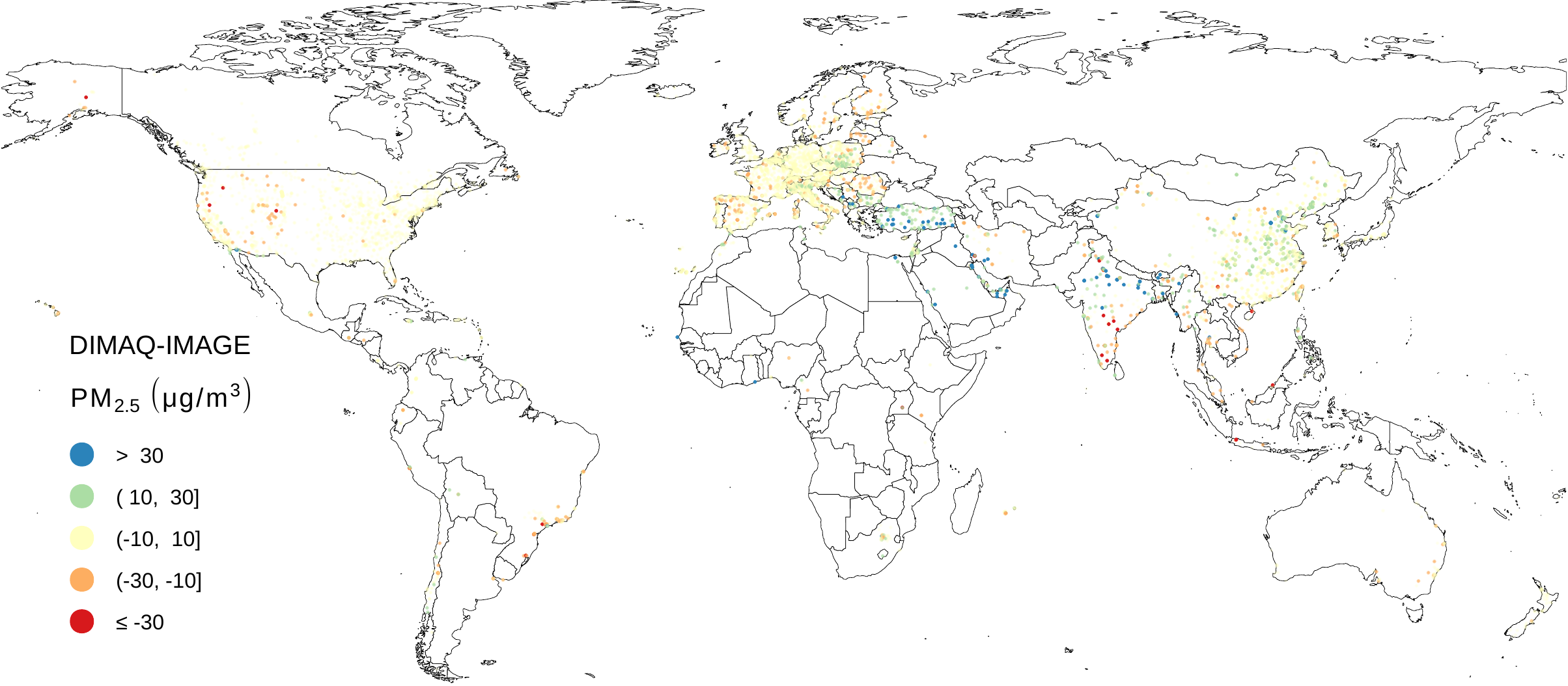}
\end{center}
\caption{Differences in predicted long-term average PM$_{2.5}$ concentrations (2010-2016) using the IMAGE-PM$_{2.5}$ model and the DIMAQ model~\cite{shaddick2018data,shaddick2018datab}.}
\label{fig:comp_dimaq_map}
\end{figure*}

Continuous estimates of annual average PM$_{2.5}$ concentrations from our global IMAGE-PM$_{2.5}$ model were highly correlated (R$^2$=0.79; slope = 1.019, 95\% CI: 1.014, 1.025) with those predicted by the Data Integration Model for Air Quality (DIMAQ) used by the Global Burden of Disease (GBD) study (Figure~\ref{fig:comp_dimaq}). Agreement between the two models improved slightly when we compared our global IMAGE-PM$_{2.5}$ predictions to DIMAQ model estimates averaged over the entire seven-year period tested (2010-2016): R$^2$=0.81; slope=1.022 (95\% CI: 1.012, 1.025). Figure~\ref{fig:comp_dimaq_map} shows the global distribution of differences between long-term estimates of mean PM$_{2.5}$ concentrations (2010-2016) at the 9000 sites compared in this analysis. Agreement was best in North America, Europe, and China. The largest differences were observed in regions where ground level PM$_{2.5}$ values (used in DIMAQ) were based predominantly ($>$70\% of values) on PM$_{10}$data including India, Turkey, Romania, and Lithuania. 

\section{Discussion}\label{sec:discussion}
In this study we explored the use of deep convolutional neural networks as an alternative, cost-effective means of estimating global variations in long-term average outdoor PM$_{2.5}$ concentrations. In particular, we examined this approach across the global concentration range using ground monitoring data available from the WHO as well as across the more limited concentration range in North America using PM$_{2.5}$ predictions based on remote sensing~\cite{van2016global}. To our knowledge, this is the first study to explore the use of deep convolutional neural networks in estimating global variations in annual average outdoor PM$_{2.5}$ concentrations and we noted several interesting findings.

First, the predictive performance of the global IMAGE-PM$_{2.5}$ model presented in this study was similar to that of current state-of-the-art Bayesian hierarchical models employing combinations of remote sensing, chemical transport models, land use, and other information~\cite{shaddick2018data,shaddick2018datab}. This is somewhat surprising given the wealth of source/emissions information included in state-of-the-art models. Specifically, Shaddick~\emph{et al.}~\cite{shaddick2018data,shaddick2018datab} reported a population-weighted RMSE value of 12.10 $\mu$g/m$^3$ (R$^2$=0.91) for the DIMAQ model used in the Global Burden of Disease Study whereas the IMAGE-PM$_{2.5}$ in our investigation achieved an RMSE value of 13.01 $\mu$g/m$^3$ (R$^2$=0.75) over a similar concentration range. In addition, our direction comparison of DIMAQ and IMAGE-PM$_{2.5}$ predictions indicated a strong correlation between model estimates with a slope close to 1. Interestingly, the largest discrepancies between the two models occurred in regions where ground level PM$_{2.5}$ data were derived from PM$_{10}$ measurements. As the DIMAQ model incorporated PM$_{2.5}$ data derived from PM10 measurements and the IMAGE-PM$_{2.5}$ model did not, this difference may explain the larger discrepancies in these areas.

The North American model presented here is not directly comparable to values reported by Shaddick~\emph{et al.}~\cite{shaddick2018data,shaddick2018datab} because it covers a narrow exposure range and is in fact a model of modelled values~\cite{van2016global,atmcomp}.  Specifically, the total error in our North American model compared to ground measurements would be the sum of errors in remote sensing estimates (compared to measured ground-level PM$_{2.5}$ concentrations) plus the additional error contributed by our model. Nevertheless, our findings from North America are important in that they suggest that deep convolutional neural networks may be used to estimate spatial variations in long-term average PM$_{2.5}$ concentrations over a narrow range of concentrations. Moreover, our findings indicate that deep learning model estimates based on satellite images may offer an additional source of information for state-of-the-art Bayesian hierarchical models such as DIMAQ~\cite{shaddick2018data,shaddick2018datab} which integrate multiple complex data streams. In particular, our IMAGE-PM$_{2.5}$ model may offer useful prior information for Bayesian models when ground level measurements or emissions data are not available.  

One of the clear disadvantages of deep learning models is the lack of transparency in how model predictions are generated. Deep convolutional neural networks are somewhat less opaque in that class activation maps and filter visualizations can be used to investigate image characteristics/patterns used to make predictions. Our results suggest that model predictions of ground-level PM$_{2.5}$ concentrations were based on localized portions of satellite images and that both color and combinations of colors and geometric features (\ie lines/edges) were used in making predictions. However, it was not possible to identify specific aspects of the built environment that played an important role in generating model estimates. Interestingly, the zoom level of satellite images had an important impact on model performance and future studies should explore other image characteristics that could be optimized to reduce model errors. Likewise, as deep convolutional neural networks can have multiple inputs, it may be possible to incorporate additional ground-level information (\eg sources, businesses, population density, etc.) within each image to capture more detailed data on local sources of PM$_{2.5}$ and thus improve model performance.  	 

A second limitation of our analysis was that the timing of satellite images did not overlap exactly with the timing of PM$_{2.5}$ measurements/estimates. This may have contributed to error to our predictions if major infrastructure changes were made between the time of PM$_{2.5}$ measurements and satellite imaging.  Moreover, our IMAGE-PM$_{2.5}$ model is also limited in that it does not contain a temporal component: predictions only change if the image changes. Therefore, the IMAGE-PM$_{2.5}$ model cannot be used to estimate short-term (\ie year to year) changes in outdoor PM$_{2.5}$ concentrations and this limitation will be addressed in our ongoing work.

In summary, we developed a new method of estimating global variations in long-term average outdoor PM$_{2.5}$ concentrations using deep convolutional neural networks trained with large databases of satellite images and ground level measurements. Our new global IMAGE-PM$_{2.5}$ model relies on a single input (a satellite image) and can provide fast, cost-effective estimates of PM$_{2.5}$ concentrations with predictive performance comparable to modern Bayesian hierarchical models currently used by the Global Burden of Disease Project~\cite{shaddick2018data,shaddick2018datab}. These findings represent an important advance in our current understanding of how global variations in long-term average PM$_{2.5}$ concentrations can be modelled for global health applications. The IMAGE-PM$_{2.5}$ model can be used as a stand-alone method of global exposure estimation or incorporated into more complex hierarchical model structures.

{\small
\bibliographystyle{ieee_fullname}
\bibliography{bibliography}

\begin{thebibliography}{10}\itemsep=-1pt

\bibitem{atmcomp}
Atmospheric composition analysis group. satellite-derived pm2.5 with gwr, north
  american, 2010-2012, at 35\% rh [ug/m3].
\newblock \url{http://fizz.phys.dal.ca/~atmos/martin/?page_id=140}.
\newblock Accessed: 2019-03-24.

\bibitem{angermueller2016deep}
Christof Angermueller, Tanel P{\"a}rnamaa, Leopold Parts, and Oliver Stegle.
\newblock Deep learning for computational biology.
\newblock {\em Molecular systems biology}, 2016.

\bibitem{brauer2012exposure}
Michael Brauer, Markus Amann, Rick~T Burnett, Aaron Cohen, Frank Dentener,
  Majid Ezzati, Sarah~B Henderson, Michal Krzyzanowski, et~al.
\newblock Exposure assessment for estimation of the global burden of disease
  attributable to outdoor air pollution.
\newblock {\em Environmental science \& technology}, 2012.

\bibitem{bridle90softmax}
John Bridle.
\newblock Probabilistic interpretation of feedforward classification network
  outputs, with relationships to statistical pattern recognition.
\newblock {\em Neurocomputing: Algorithms, Architectures and Applications},
  1990.

\bibitem{keras}
Fran{\c{c}}ois Chollet.
\newblock Keras.
\newblock \url{https://keras.io}.
\newblock Accessed: 2019-03-24.

\bibitem{chollet2017xception}
Fran{\c{c}}ois Chollet.
\newblock Xception: Deep learning with depthwise separable convolutions.
\newblock In {\em Proceedings of the IEEE conference on computer vision and
  pattern recognition}, 2017.

\bibitem{cruz2017accurate}
Angel Cruz-Roa, Hannah Gilmore, Ajay Basavanhally, Michael Feldman, Shridar
  Ganesan, Natalie~NC Shih, John Tomaszewski, Fabio~A Gonz{\'a}lez, and Anant
  Madabhushi.
\newblock Accurate and reproducible invasive breast cancer detection in
  whole-slide images: A deep learning approach for quantifying tumor extent.
\newblock {\em Scientific reports}, 2017.

\bibitem{dozat2016nadam}
Timothy Dozat.
\newblock Incorporating nesterov momentum into adam.
\newblock 2016.

\bibitem{esteva2017dermatologist}
Andre Esteva, Brett Kuprel, Roberto~A Novoa, Justin Ko, Susan~M Swetter,
  Helen~M Blau, and Sebastian Thrun.
\newblock Dermatologist-level classification of skin cancer with deep neural
  networks.
\newblock {\em Nature}, 2017.

\bibitem{gulshan2016development}
Varun Gulshan, Lily Peng, Marc Coram, Martin~C Stumpe, Derek Wu, Arunachalam
  Narayanaswamy, Subhashini Venugopalan, Kasumi Widner, Tom Madams, Jorge
  Cuadros, et~al.
\newblock Development and validation of a deep learning algorithm for detection
  of diabetic retinopathy in retinal fundus photographs.
\newblock {\em Jama}, 2016.

\bibitem{ggmap}
David Kahle and Hadley Wickham.
\newblock ggmap: Spatial visualization with ggplot2.
\newblock {\em The R Journal}, 2013.

\bibitem{landrigan2018lancet}
Philip~J Landrigan, Richard Fuller, Nereus~JR Acosta, Olusoji Adeyi, et~al.
\newblock The lancet commission on pollution and health.
\newblock {\em The Lancet}, 2018.

\bibitem{lecun1998gradient}
Yann LeCun, L{\'e}on Bottou, Yoshua Bengio, Patrick Haffner, et~al.
\newblock Gradient-based learning applied to document recognition.
\newblock {\em Proceedings of the IEEE}, 1998.

\bibitem{maharana2018use}
Adyasha Maharana and Elaine~Okanyene Nsoesie.
\newblock Use of deep learning to examine the association of the built
  environment with prevalence of neighborhood adult obesity.
\newblock {\em JAMA}, 2018.

\bibitem{geohash}
G. Niemeyer.
\newblock Geohash.
\newblock \url{http://geohash.org}.
\newblock Accessed: 2019-03-24.

\bibitem{who16data}
World~Health Organization.
\newblock Who global urban ambient air pollution database (update 2016).
\newblock \url{https://whoairquality.shinyapps.io/AmbientAirQualityDatabase/}.
\newblock Accessed: 2019-03-24.

\bibitem{patton2015transferability}
Allison~P Patton, Wig Zamore, Elena~N Naumova, Jonathan~I Levy, Doug Brugge,
  and John~L Durant.
\newblock Transferability and generalizability of regression models of
  ultrafine particles in urban neighborhoods in the boston area.
\newblock {\em Environmental science \& technology}, 2015.

\bibitem{ryan2007review}
Patrick~H Ryan and Grace~K LeMasters.
\newblock A review of land-use regression models for characterizing intraurban
  air pollution exposure.
\newblock {\em Inhalation toxicology}, 2007.

\bibitem{selvaraju2017grad}
Ramprasaath~R Selvaraju, Michael Cogswell, Abhishek Das, Ramakrishna Vedantam,
  Devi Parikh, and Dhruv Batra.
\newblock Grad-cam: Visual explanations from deep networks via gradient-based
  localization.
\newblock In {\em CVPR}, 2017.

\bibitem{shaddick2018data}
Gavin Shaddick, Matthew~L Thomas, Heresh Amini, David Broday, Aaron Cohen,
  Joseph Frostad, Amelia Green, Sophie Gumy, Yang Liu, Randall~V Martin, et~al.
\newblock Data integration for the assessment of population exposure to ambient
  air pollution for global burden of disease assessment.
\newblock {\em Environmental science \& technology}, 2018.

\bibitem{shaddick2018datab}
Gavin Shaddick, Matthew~L Thomas, Amelia Green, Michael Brauer, Aaron
  Donkelaar, Rick Burnett, Howard~H Chang, Aaron Cohen, Rita~Van Dingenen,
  Carlos Dora, et~al.
\newblock Data integration model for air quality: a hierarchical approach to
  the global estimation of exposures to ambient air pollution.
\newblock {\em Journal of the Royal Statistical Society: Series C (Applied
  Statistics)}, 2018.

\bibitem{simonyan2014vgg}
Karen Simonyan and Andrew Zisserman.
\newblock Very deep convolutional networks for large-scale image recognition.
\newblock {\em ICLR}, 2015.

\bibitem{stanaway2018global}
Jeffrey~D Stanaway, Ashkan Afshin, Emmanuela Gakidou, Stephen~S Lim, Degu
  Abate, , et~al.
\newblock Global, regional, and national comparative risk assessment of 84
  behavioural, environmental and occupational, and metabolic risks or clusters
  of risks for 195 countries and territories, 1990--2017: a systematic analysis
  for the global burden of disease study 2017.
\newblock {\em The Lancet}, 2018.

\bibitem{szegedy2016rethinking}
Christian Szegedy, Vincent Vanhoucke, Sergey Ioffe, Jon Shlens, and Zbigniew
  Wojna.
\newblock Rethinking the inception architecture for computer vision.
\newblock In {\em CVPR}, 2016.

\bibitem{Rlang}
R~Development~Core Team.
\newblock {\em R: A language and environment for statistical computing}.
\newblock R Foundation for Statistical Computing, 2010.

\bibitem{tieleman2012rmsprop}
T. Tieleman and G. Hinton.
\newblock {Lecture 6.5---RmsProp: Divide the gradient by a running average of
  its recent magnitude}.
\newblock COURSERA: Neural Networks for Machine Learning, 2012.

\bibitem{van2016global}
Aaron Van~Donkelaar, Randall~V Martin, Michael Brauer, N~Christina Hsu, Ralph~A
  Kahn, Robert~C Levy, Alexei Lyapustin, Andrew~M Sayer, and David~M Winker.
\newblock Global estimates of fine particulate matter using a combined
  geophysical-statistical method with information from satellites, models, and
  monitors.
\newblock {\em Environmental science \& technology}, 2016.

\bibitem{weichenthal2019picture}
Scott Weichenthal, Marianne Hatzopoulou, and Michael Brauer.
\newblock A picture tells a thousand… exposures: Opportunities and challenges
  of deep learning image analyses in exposure science and environmental
  epidemiology.
\newblock {\em Environment international}, 2019.

\bibitem{weichenthal2016characterizing}
Scott Weichenthal, Keith Van~Ryswyk, Alon Goldstein, Maryam Shekarrizfard, and
  Marianne Hatzopoulou.
\newblock Characterizing the spatial distribution of ambient ultrafine
  particles in toronto, canada: A land use regression model.
\newblock {\em Environmental pollution}, 2016.

\end{thebibliography}
}

\end{document}